\documentclass[10pt]{amsart}
\usepackage{amssymb}
\usepackage{amsmath,amssymb,latexsym,amsfonts,amsthm,
latexsym, amscd, amsfonts, epsfig,epic,psfrag,eepic}
\usepackage{mathrsfs}
\usepackage{color}
\usepackage{eucal}%    caligraphic-euler fonts: \mathcal{ }
\usepackage{eufrak}%   frak-euler        fonts: \mathfrak{ }
\usepackage[all]{xypic}
\usepackage{xspace}
\usepackage{array}
\usepackage{graphicx}
\theoremstyle{plain}
\newtheorem{theorem}{Theorem}

\newtheorem{lemma}{Lemma}

\theoremstyle{definition}

\theoremstyle{example}

\theoremstyle{remark}

\numberwithin{equation}{section}

\begin{document}

\title[On the decomposition of $k$-noncrossing RNA structures]
      {On the decomposition of $k$-noncrossing RNA structures}
\author{Emma Y. Jin and Christian M. Reidys$^{\,\star}$}
\address{Center for Combinatorics, LPMC-TJKLC %XXX%
           \\
         Nankai University  \\
         Tianjin 300071\\
         P.R.~China\\
         Phone: *86-22-2350-6800\\
         Fax:   *86-22-2350-9272}
\email{reidys@nankai.edu.cn}
\thanks{}
\date{January, 2009}
\begin{abstract}
An $k$-noncrossing RNA structure can be identified with an
$k$-noncrossing diagram over $[n]$, which in turn corresponds to a
vacillating tableaux having at most $(k-1)$ rows. In this paper we
derive the limit distribution of irreducible substructures via
studying their corresponding vacillating tableaux. Our main
result proves, that the limit distribution of the numbers of irreducible
substructures in $k$-noncrossing, $\sigma$-canonical RNA structures is
determined by the density function of a $\Gamma(-\ln\tau_k,2)$-distribution
for some $\tau_k<1$.
\end{abstract}
\maketitle {{\small
%\tableofcontents
}}
%%%
%%%
%%%%%%%%%%%%%%%%%%%%%%%%%%%%%%%%%%%%%%%%%%%%%%%%%%%%%%%%%%%%%%%%%%%%%%%%
%%%
%%%
\section{Introduction and background}

%%%
%%%%%%%%%%%%%%%%%%%%%%%%%%%%%%%%%%%%%%%%%%%%%%%%%%%%%%%%%%%%%%%%%%%%%%%%
%%%

In this paper we analyze the number of irreducible substructures of
$k$-noncrossing, $\sigma$-canonical RNA structures. We prove that
the numbers of irreducible substructures of $k$-noncrossing,
$\sigma$-canonical RNA structures are, in the limit of long
sequence length, given via the density function of a
$\Gamma(-\ln\tau_k,2)$-distribution.

An RNA structure is the helical configuration of its primary
sequence, i.e.~the sequence of nucleotides {\bf A}, {\bf G}, {\bf U}
and {\bf C}, together with Watson-Crick ({\bf A-U}, {\bf G-C}) and
({\bf U-G}) base pairs. As RNA structure is oftentimes tantamount to
its function, it is of key importance. The concept of irreducibility
in RNA structures is of central importance since the computation of
the minimum free energy (mfe) configuration of a given RNA molecule
is determined by its largest, irreducible substructure.

\begin{figure}[ht]
\centerline{%
\epsfig{file=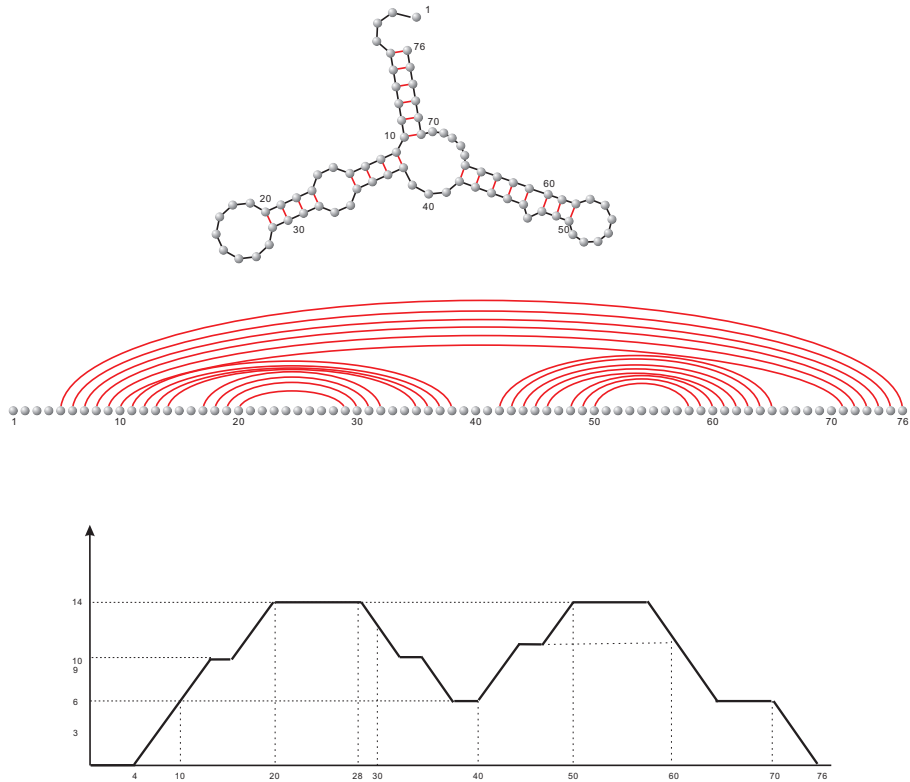,width=0.8\textwidth}\hskip15pt
 }
\caption{\small The phenylalanine tRNA secondary structure, as generated
by the computer folding algorithm {\sf cross} \cite{Reidys:08algo},
represented as planar graph, diagram
and Motzkin-path.
The structure has arc-length $\ge 8$ and stack-length $\ge 3$ and
uniquely corresponds to a Motzkin-path with minimum height $3$ and minimum
plateau-length $7$.} \label{F:sec}
\end{figure}

Three decades ago, Waterman
\cite{Penner:93c,Waterman:79a,Waterman:78a,Waterman:80,Waterman:94a}
pioneered the combinatorics of RNA secondary structures, an RNA
structure class exhibiting only noncrossing bonds. Secondary
structures can readily be identified with Motzkin-paths satisfying
some minimum height and plateau-length, see Figure~\ref{F:sec}.
%%%
%%%%%%%%%%%%%%%%%%%%%%%%%%%%%%%%%%%%%%%%%%%%%%%%%%%%%%%%%%%%%%%%%%%%%%%%
%%%
%%%
%%%%%%%%%%%%%%%%%%%%%%%%%%%%%%%%%%%%%%%%%%%%%%%%%%%%%%%%%%%%%%%%%%%%%%%%
%%%
The latter restrictions arise from biophysical constraints due to
mfe and the limited flexibility of chemical bonds.
It is clear from the
particular bijection, that irreducible substructures in RNA secondary
structures are closely related to the number of nontrivial returns, i.e.~the
number of non-endpoints, for which the Motzkin-path meets the $x$-axis.

For Dyck-paths this question has been studied by Shapiro \cite{Shapiro},
who showed that the expected number of nontrivial returns of Dyck-paths of
length $2n$ equals $\frac{2n-2}{n+2}$. Subsequently, Shapiro and Cameron
\cite{Cameron:03}
derived expectation and variance of the number of nontrivial returns for
generalized Dyck-paths from $(0,0)$ to $((t+1)n,0)$
\begin{equation}\label{E:formel}
\mathbb{E}[\xi_t]=\frac{2n-2}{tn+2}\quad \mbox { and }\quad
\mathbb{V}[\xi_t]=\frac{2tn(n-1)((t+1)n+1)}{(tn+2)^2(tn+3)}.
\end{equation}
The bijection between Dyck-path of length $2n$
and the unique triangulation of the $(n+2)$-gon, due to Stanley
\cite{Stanley:99}, implies a combinatorial proof for $\mathbb{E}[\xi_1]$.
An alternative approach is to employ the Riordan matrix \cite{Shapiro:91},
an infinite, lower triangular matrix $L=(l_{n,k})_{n,k\ge 0}=(g,f)$,
where $g(z)=\sum_{n \ge 0}g_n z^n$, $f(z)=\sum_{n \ge 0}f_nz^n$ with
$f_0=0,f_1\ne 0$, such that $\sum_{n \ge k}l_{n,k}z^n=g(z)f^k(z)$.
Clearly,
\begin{equation*}
{\bf C}(z)=\sum_{n\ge 0}C_n z^n=\frac{1-\sqrt{1-4z}}{2z} \mbox{\quad
where } C_n=\frac{1}{n+1}{2n\choose n}
\end{equation*}
is the generating function of Dyck-paths and let $\zeta_{n,j}$
denote the number of Dyck-paths of length $2n$ with $j$ nontrivial
returns. We consider the Riordan matrix $L=(\zeta_{n,j})_{n,j\ge
0}=(z{\bf C}(z),z{\bf C}(z))$ and extract the coefficients
$\zeta_{n,j}$ from its generating function $(z{\bf C}(z))^{j+1}$ by
Lagrange inversion. Setting $f(z)=zG(f(z))$ with $f(z)={\bf C}(z)-1$
and $G(z)=(1+z)^2$, we obtain
\begin{equation*}
\zeta_{n,j} =[z^{n-j-1}](f(z)+1)^{j+1} = \frac{j+1}{2n-j-1}{2n-j-1
\choose n},
\end{equation*}
where $\sum_{j\ge 0}\zeta_{n,j}=C_n$. From this we immediately
compute $\mathbb{E}[\xi_{1}]  =  \sum_{j\ge 1}j\cdot
\frac{\zeta_{n,j}}{C_n}$ and $\mathbb{V}[\xi_1]  =  \sum_{j\ge 1}j^2
\cdot \frac{\zeta_{n,j}}{C_n}-\left(\sum_{j\ge 1}j\cdot
\frac{\zeta_{n,j}}{C_n}\right)^2$, from which the expression of
eq.~(\ref{E:formel}), for $t=1$ follows.

In Section~\ref{S:gap} we consider the bivariate generating function
directly, which relates to the Riordan matrix in case of generalized Dyck-path
as follows
\begin{equation*}
\sum_{n \ge 0}\sum_{j \ge 0}\zeta_{n,j}w^jz^n = \sum_{j \ge
0}z^{j+1}{\bf C}(z)^{j+1}w^j = \frac{z{\bf C}(z)}{1-wz{\bf C}(z)}.
\end{equation*}
Our main idea is to derive the bivariate generating function from the
Riordan matrix employing irreducible paths and to establish via
singularity analysis a discrete
limit law. This is done, however, for the far more general class of
$\mathfrak{C}$-tableaux introduced in Section~\ref{S:back}: in
Theorem~\ref{T:central} we show that the limit distribution of
nontrivial returns for these vacillating tableaux is given in terms of
 the density
function of a $\Gamma(\lambda,r)$-distribution, which is, already
for Motzkin-paths, a new result. For restricted Motzkin-paths
satisfying specific height and plateau-lengths, the Riordan matrix
Ansatz does not work ``directly'', since the inductive decomposition
of restricted Motzkin-paths is incompatible. Instead we introduce
the notion of irreducible paths and express the Riordan matrix in
terms of the latter, see Lemma~\ref{L:bistar}. This Ansatz allows us
to compute the generating function of irreducible paths via setting
one indeterminate of the bivariate generating function to one. The
framework developed in Section~\ref{S:gap} and
Section~\ref{S:empty}, in fact works as long as the generating
function of the particular path-class has a singular expansion and
is explicitly known. We have, for instance, for nontrivial returns
of Motzkin-paths with height $\ge 3$ and plateau length $\ge 3$:
$\lim_{n\rightarrow\infty} \mathbb{E}[\eta_n]\approx 0.8625$ and
$\lim_{n\rightarrow\infty}\mathbb{V}[\eta_n]\approx 1.2343$.

\begin{figure}[ht]
\centerline{%
\epsfig{file=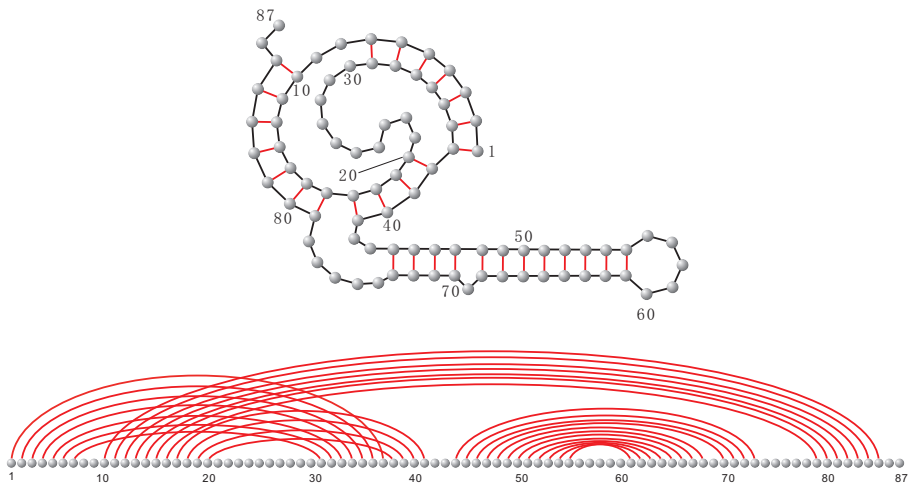,width=0.80\textwidth}\hskip15pt
 }
\caption{\small The hepatitis delta virus (HDV)-pseudoknot structure
and its diagram representation. Top: the structure as
folded by {\sf cross} \cite{Reidys:08algo} for $k=3$ and minimum
stack size $3$ and the corresponding diagram representation (bottom).}
\label{F:reid3.eps}
\end{figure}

Indeed, RNA structures are far more complex than secondary
structures: they exhibit additional, cross-serial nucleotide
interactions \cite{Searls}. These interactions were observed in
natural RNA structures, as well as via comparative sequence analysis
\cite{Westhof:92a}. They are called pseudoknots, see
Figure~\ref{F:reid3.eps}, and widely occur in functional RNA, like
for instance, eP RNA \cite{Loria:96a} as well as ribosomal RNA
\cite{Konings:95a}. RNA pseudoknots are conserved also in the
catalytic core of group I introns. In plant viral RNAs pseudoknots
mimic tRNA structure and in vitro RNA evolution \cite{Tuerk:92}
experiments have produced families of RNA structures with pseudoknot
motifs, when binding HIV-1 reverse transcriptase.

\begin{figure}[ht]
\centerline{%
\epsfig{file=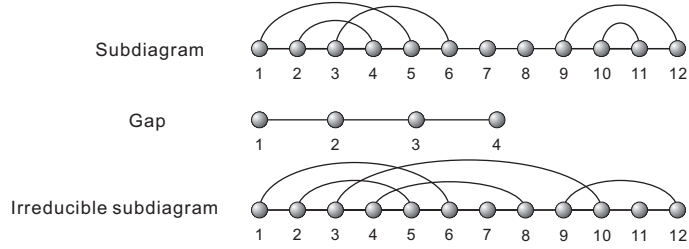,width=0.6\textwidth}\hskip15pt
 }
\caption{\small Subdiagrams, gaps and irreducibility:
a diagram (top), decomposed into the subdiagram over $(1,6)$,
the gap $(7,8)$ and the subdiagram over $(9,12)$. A gap (middle)
and an irreducible diagram over $(1,12)$.}
\label{F:gapsub}
\end{figure}

\begin{figure}[ht]
\centerline{%
\epsfig{file=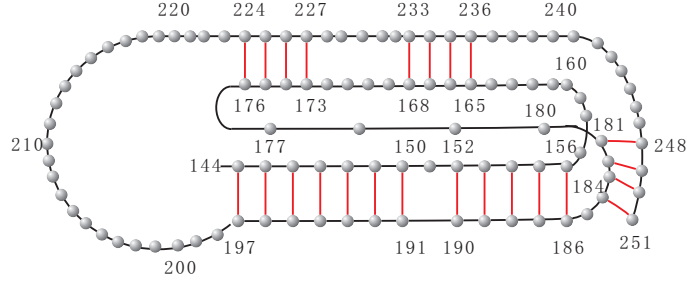,width=0.6\textwidth}\hskip15pt
 }
\caption{\small mRNA-Ec$_\alpha$: the irreducible pseudoknot structure of the
regulatory region of the $\alpha$ ribosomal protein operon.}
\label{F:pseudo1}
\end{figure}

Combinatorially, cross serial interactions are tantamount to crossing bonds.
To this end, RNA pseudoknot structures have been modeled via $k$-noncrossing
diagrams \cite{Reidys:07pseu}, i.e.~labeled graphs over the vertex set
$[n]=\{1,\dots, n\}$ with degree $\le 1$.
Diagrams are represented by drawing their vertices $1,\dots,n$
in a horizontal line and its arcs $(i,j)$, where $i<j$, in the upper
half plane. Here the degree of $i$ refers to the number of non-horizontal
arcs incident to $i$, i.e.~the backbone of the primary sequence is not
accounted for. The vertices and arcs correspond to nucleotides and
Watson-Crick ({\bf A-U}, {\bf G-C}) and ({\bf U-G}) base pairs,
respectively, see Figure~\ref{F:reid3.eps}.
%%%%%%%%%%%%%%%%%%%%%%%%%%%%%%%%%%%%%%
\begin{figure}[ht]
\centerline{%
\epsfig{file=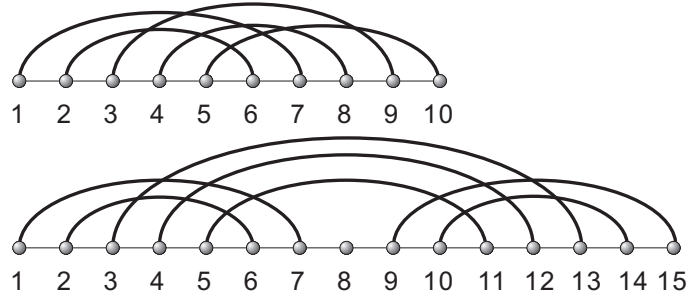,width=0.6\textwidth}\hskip15pt
 }
\caption{\small $k$-noncrossing diagrams: we display a
$4$-noncrossing, arc-length $\lambda\ge 4$ and $\sigma\ge 1$
diagram (top), where the edge set $\{(1,7),(3,9),(5,10)\}$ is a
$3$-crossing, the arc $(2,6)$ has length $4$ and $(5,10)$ has
stack-length $1$. Below, we display a $3$-noncrossing, $\lambda\ge
4$ and $\sigma\ge 2$ (lower) diagram, where $(2,6)$ has arc-length $4$
and the stack $((2,6),(1,7))$ has stack-length $2$.}
\label{F:dia}
\end{figure}
%%%%%%%%%%%%%%%%%%%%%%%%%%%%%%%%%%%%%%%%%
Diagrams are characterized via their maximum number of mutually
crossing arcs, $k-1$, their  minimum arc-length, $\lambda$, and their
minimum stack-length, $\sigma$.
A $k$-crossing is a set of $k$ distinct arcs
$(i_{1},j_{1}),(i_{2},j_{2}),\ldots(i_{k},j_{k})$
with the property $i_{1}<i_{2}<\ldots<i_{k}<j_{1}<j_{2}<\ldots<j_{k}$. A
diagram without any $k$-crossings is called a $k$-noncrossing diagram.
The length of an arc $(i,j)$ is $j-i$ and a stack of length $\sigma$
is a sequence of ``parallel'' arcs of the form
\begin{equation*}
((i,j),(i+1,j-1),\dots,(i+(\sigma-1),j-(\sigma-1))).
\end{equation*}
A subdiagram of a $k$-noncrossing diagram is a subgraph over a
subset $M\subset [n]$ of consecutive vertices that starts with an
origin and ends with a terminus of some arc.
Let $(i_1,\dots,i_m)$ be a sequence of isolated points, and $(j_1,j_2)$
be an arc. We call $(i_1,\dots,i_m)$ interior if and only if there exists
some arc $(j_1,j_2)$ such that $j_1<i_1<i_m<j_2$ holds and exterior, otherwise.
Any exterior sequence of consecutive, isolated vertices is called a gap.
A diagram or subdiagram is called irreducible, if it cannot be decomposed
into a sequence of gaps and subdiagrams, see Figure~\ref{F:gapsub}.
Accordingly, any $k$-noncrossing diagram can be uniquely decomposed into
an alternating sequence of gaps and irreducible subdiagrams. In fact
irreducibility is quite common for natural RNA pseudoknot structures, see
Figure~\ref{F:pseudo1}.
%%%%%%%%%%%%%%%
%%%%%%%%%%%%%%%%%%%%%%%%

We call a $k$-noncrossing, $\sigma$-canonical diagram with
arc-length $\ge 4$ and stack-length $\ge \sigma$, a $k$-noncrossing,
$\sigma$-canonical RNA structure, see Figure~\ref{F:dia}. We
accordingly adopt the notions of gap, substructure and
irreducibility for RNA structures.

\begin{figure}[ht]
\centerline{%
\epsfig{file=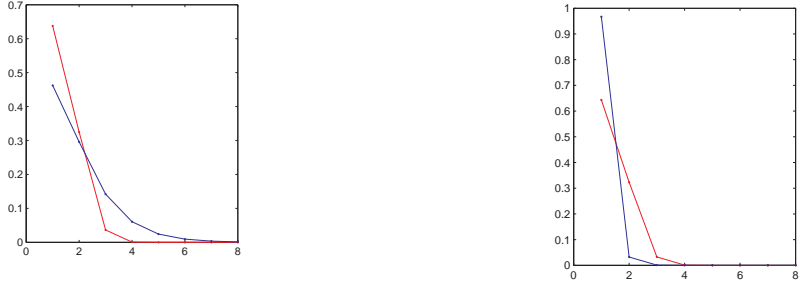,width=0.9\textwidth}\hskip15pt
 }
\caption{\small For $n=75$ the lhs displays the distribution of
irreducible substructures obtained by folding $10^4$ random
sequences into their RNA secondary structures \cite{Vienna} (red),
and the scaled density function of a
$\Gamma(-\ln(0.2241),2)$-distribution (blue) sampled at the positive
integers. The rhs shows this distribution obtained by folding
$9\times 10^3$ random sequences into $3$-noncrossing, $3$-canonical
structures \cite{Reidys:08algo} (red) and the scaled density
function of a $\Gamma(-\ln(0.0167),2)$-distribution (blue) derived
from Theorem~\ref{T:centra2}.} \label{F:experiment}
\end{figure}

Our main result is Theorem~\ref{T:centra2}, which proves that the
numbers of irreducible substructures are in the limit of long
sequence length given via the density function of a
$\Gamma(-\ln\tau_k,2)$-distribution. Furthermore, we show that the
probability generating function of the limit distribution is given
by $q(u)=\frac{u(1-\tau_k)^2}{(1-\tau_k u)^2}$, where $\tau_k$ is
expressed in terms of the generating function of $k$-noncrossing,
$\sigma$-canonical RNA structures \cite{Reidys:08ma} and its
dominant singularity $\alpha_k$. In Figure~\ref{F:experiment} we
compare our analytic results with mfe secondary and $3$-noncrossing
structures generated by computer folding algorithms
\cite{Vienna,Reidys:08algo}, respectively. The data indicate that
already for $n=75$, the limit distribution of
Theorem~\ref{T:centra2} provides for both structure classes a good
fit.

%%%
%%%%%%%%%%%%%%%%%%%%%%%%%%%%%%%%%%%%%%%%%%%%%%%%%%%%%%%%%%%%%%%%%%%%%%%%%%
%%%
%%%
%%%%%%%%%%%%%%%%%%%%%%%%%%%%%%%%%%%%%%%%%%%%%%%%%%%%%%%%%%%%%%%%%%%%%%%%%%
%%%

The paper is organized as follows: in Section~\ref{S:back} we recall
some basic combinatorial background. Of particular importance here
is the bijection between $k$-noncrossing diagrams and vacillating
tableaux of Theorem~\ref{T:cross} with at most $(k-1)$ rows
\cite{Reidys:07vac}.
In Section~\ref{S:gap}, we present all key ideas and derive the limit
distribution of $*$-tableaux. In Section~\ref{S:empty} we study the
limit distribution of nontrivial returns using the framework developed in
Section~\ref{S:gap}.

%%%
%%%%%%%%%%%%%%%%%%%%%%%%%%%%%%%%%%%%%%%%%%%%%%%%%%%%%%%%%%%%%%%%%%%%%%
%%%
\section{Some basic facts}\label{S:back}
%%%
%%%%%%%%%%%%%%%%%%%%%%%%%%%%%%%%%%%%%%%%%%%%%%%%%%%%%%%%%%%%%%%%%%%%%%
%%%

\begin{figure}[ht]
\centerline{\epsfig{file=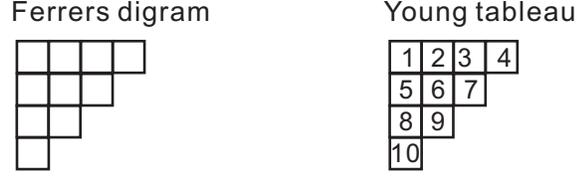,width=0.5\textwidth} \hskip8pt}
\caption{\small Ferrers diagram and Young tableau.}\label{F:fyt}
\end{figure}

\begin{figure}[ht]
\centerline{\epsfig{file=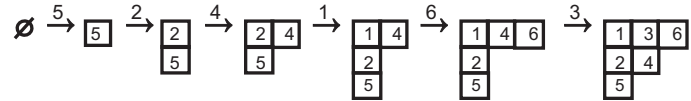,width=0.6\textwidth} \hskip8pt}
\caption{\small RSK-insertion of the elements $5,2,4,1,6,3$. The insertion
of the above sequence successively constructs a standard Young tableau.
}\label{F:tanrsk}
\end{figure}

A \emph{Ferrers diagram} (shape) is a collection of squares arranged in
left-justified rows with weakly decreasing number of boxes in each row.
A standard \emph{Young tableau} (SYT) is a filling of the squares by numbers
which is strictly decreasing in each row and in each column. We refer to
standard Young tableaux as Young tableaux, see Figure~{\ref{F:fyt}}.
%%%
%%%%%%%%%%%%%%%%%%%%%%%%%%%%%%%%%%%%%%%%%%%%%%%%%%%%%%%%%%%%%%%%%%%%%%%%
%%%%%%%%%%%%%%%%%%%%%%%%%%%%%%%%%%%%%%%%%%%%%%%%%%%%%%%%%%%%%%%%%%%%%%%%%
%%%
A \emph{vacillating tableau} $V_{\lambda}^{2n}$ of shape $\lambda$ and length
$2n$ is a sequence of Ferrers diagrams
$
(\lambda^{0}, \lambda^{1},\ldots, \lambda^{2n})
$
of shapes such that { (i)} $\lambda^{0}=\varnothing$
and $\lambda^{2n}=\lambda,$ and { (ii)}
$(\lambda^{2i-1},\lambda^{2i})$ is derived from
           $\lambda^{2i-2}$, for $1\le i\le n$, by one of the following
operations. $(\varnothing,\varnothing)$: do nothing twice;
$(-\square,\varnothing)$: first remove a square then do nothing;
$(\varnothing,+\square)$: first do nothing then adding a square;
$(\pm \square,\pm \square)$: add/remove a square at the odd and even
steps, respectively. We denote the set of vacillating tableaux by
$\mathcal{V}_\lambda^{2n}$.
%%%
%%%%%%%%%%%%%%%%%%%%%%%%%%%%%%%%%%%%%%%%%%%%%%%%%%%%%%%%%%%%%%%%%%%%%%%%
\begin{figure}[ht]
\centerline{\epsfig{file=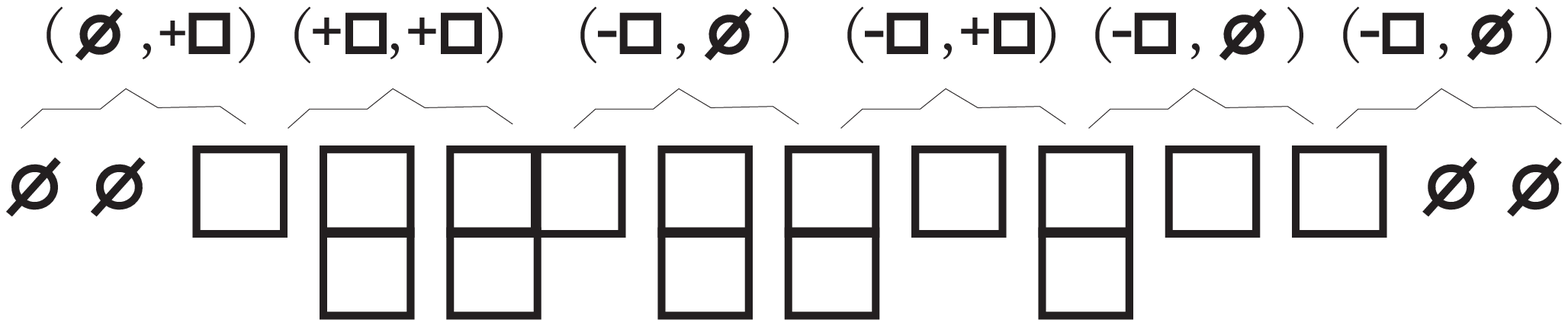,width=0.5\textwidth} \hskip8pt}
\caption{\small A vacillating tableau of shape
$\varnothing$ and length $12$.}\label{F:tanvt1}
\end{figure}
%%%%%%%%%%%%%%%%%%%%%%%%%%%%%%%%%%%%%%%%%%%%%%%%%%%%%%%%%%%%%%%%%%%%%%%%%
%%%
The \emph{RSK-algorithm} is a process of row-inserting elements into a
Young tableau.
Suppose we want to insert $q$ into a standard Young tableau of shape
$\lambda$.
Let $\lambda_{i,j}$ denote the element in the $i$-th row
and $j$-th column of the Young tableau. Let $j$ be the largest
integer such that $\lambda_{1,j-1}\le q$. (If $\lambda_{1,1}>q$,
then $j=1$.) If $\lambda_{1,j}$ does not exist, then simply add $q$
at the end of the first row. Otherwise, if $\lambda_{1,j}$ exists,
then replace $\lambda_{1,j}$ by $q$. Next insert $\lambda_{1,j}$
into the second row following the above procedure and continue until
an element is inserted at the end of a row. As a result, we obtain a
new standard Young tableau with $q$ included. For instance, inserting
the sequence $5,2,4,1,6,3$, starting with an empty shape yields the
standard Young tableaux displayed in Figure~{\ref{F:tanrsk}}.\\
%%%
%%%%%%%%%%%%%%%%%%%%%%%%%%%%%%%%%%%%%%%%%%%%%%%%%%%%%%%%%%%%%%%%%%%%%%%%
%%%%%%%%%%%%%%%%%%%%%%%%%%%%%%%%%%%%%%%%%%%%%%%%%%%%%%%%%%%%%%%%%%%%%%
%%%
The RSK-insertion algorithm has an inverse \cite{Reidys:07vac},
see Lemma~\ref{L:extract} below, which will be of central importance for
constructing a vacillating tableaux from a tangled diagram.
%%%%%
%%%%%%%%%%%%%%%%%%%%%%%%%%%%%%%%%%%%%%%%%%%%%%%%%%%%%%%%%%%%%%%%%%%
%%%%%
\begin{lemma}{\label{L:extract}}
Suppose we are given two shapes $\lambda^{i}\subsetneq\lambda^{i-1}$,
which differ by exactly one square. Let $T_{i-1}$ and $T_{i}$ be SYT
of shape $\lambda^{i-1}$ and $\lambda^{i}$, respectively.
Given $\lambda^{i}$ and $T_{i-1}$, then there
exists a unique $j$
contained in $T_{i-1}$ and a unique tableau $T_{i}$ such that
$T_{i-1}$ is obtained from $T_{i}$ by inserting $j$ via the
$\text{\rm RSK}$-algorithm.
\end{lemma}
%%%%%
%%%%%%%%%%%%%%%%%%%%%%%%%%%%%%%%%%%%%%%%%%%%%%%%%%%%%%%%%%%%%%%%%%%
%%%%%

\begin{figure}[ht]
\centerline{\epsfig{file=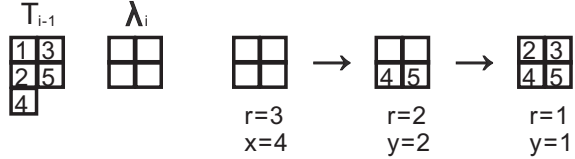,width=0.5\textwidth} \hskip8pt}
\caption{How Lemma~\ref{L:extract} works:
Given the Young tableau, $T_{i-1}$ and the shape $\lambda_{i}$,
we show how to find the unique $j$ (note here we have $j=1$)
such that $T_{i-1}$ is obtained from $T_{i}$ by inserting
$1$ via the $\text{\rm RSK}$-algorithm.
}\label{F:inverse}
\end{figure}

In addition, Lemma~\ref{L:extract} explicitly constructs this unique $j$ such
that $T_{i-1}$ is obtained from $T_{i}$ by inserting $j$ via the
$\text{\rm RSK}$-algorithm, see Figure~\ref{F:inverse}.
%%%
%%%%%%%%%%%%%%%%%%%%%%%%%%%%%%%%%%%%%%%%%%%%%%%%%%%%%%%%%%%%%%%%%%%%%%%%
%%%%%%%%%%%%%%%%%%%%%%%%%%%%%%%%%%%%%%%%%%%%%%%%%%%%%%%%%%%%%%%%%%%%
%%%

%%%%%%%%%%%%%%%%%%%%%%%%%%%%%%%%%%%%%%%%%%%%%%%%%%%%%%%%%%%%%%%%%%%
\subsection{From diagrams to vacillating tableaux and back}
\begin{figure}[ht]
\centerline{\epsfig{file=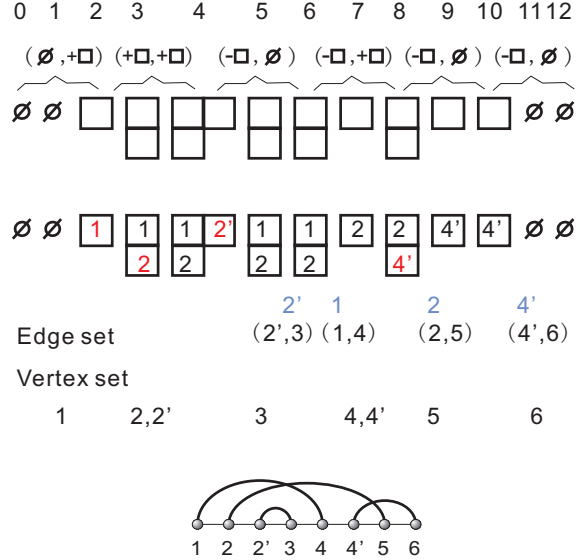,width=0.5\textwidth} \hskip8pt}
\caption{\small From vacillating tableaux to tangled diagrams:
For $+\square$ steps one simply inserts into the tableaux, does nothing
for $\varnothing$ and RSK-extracts (Lemma~\ref{L:extract}) for $-\square$.
The blue numbers $2',1,2,4'$ are obtained by RSK-extraction, corresponding
to the ``$-\square$'' steps for $i=3,4,5,6$.}\label{F:phi1}
\end{figure}

%%%%%%%%%%%%%%%%%%%%%%%%%%%%%%%%%%%%%%%%%%%%%%%%%%%%%%%%%%%%%%%%%%%
RNA tertiary interactions, in particular the interactions between
helical and non-helical regions give rise to consider tangled
diagrams \cite{Reidys:07vac}. The key feature of tangled diagrams
(tangles) is to allow for \emph{two} interactions: one being
Watson-Crick or {\bf G-U} and the other being a hydrogen bond for
each nucleotide. A \emph{tangled diagram}, $G_n$, over $[n]$ is
obtained by drawing its arcs in the upper halfplane having vertices
of degree at most two and a specific notion of crossings and
nestings \cite{Reidys:07vac}.
%%%%%%%%%%%%%%%%%%%%%%%%%%%%%%%%%%%%%%%%%%%%%%%%%%%%%%%%%%%%%%%%%%%%%%%%
\begin{figure}[ht]
\centerline{\epsfig{file=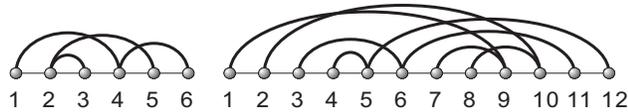,width=0.55\textwidth}
\hskip8pt} \caption{\small Tangled diagrams: the first tangled
diagram represents the key bonds of the hammerhead ribosome and the
second tangle represents key bonds of the catalytic core region of
the Group I self-splicing intron \cite{Chastain}. }\label{C08F:mmm}
\end{figure}
%%%%%%%%%%%%%%%%%%%%%%%%%%%%%%%%%%%%%%%%%%%%%%%%%%%%%%%%%%%%%%%%%%%%%%%%%%
%%%
The \emph{inflation}, of a tangle is a diagram, obtained by ``splitting''
each vertex of degree two, $j$, into two vertices $j$ and $j'$ having
degree one, see Figure~{\ref{F:inf1}}. Accordingly, a tangled diagram
with $\ell$ vertices of degree two is expanded into a diagram over
$n+\ell$ vertices. Obviously, the inflation has its unique inverse,
obtained by simply identifying the vertices $j,j'$.
%%%
%%%%%%%%%%%%%%%%%%%%%%%%%%%%%%%%%%%%%%%%%%%%%%%%%%%%%%%%%%%%%%%%%%%%%%%%
\begin{figure}[ht]
\centerline{\epsfig{file=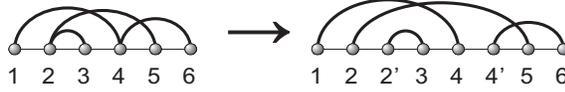,width=0.5\textwidth} \hskip8pt}
\caption{\small The inflation of the first tangled diagram in
Figure{\ref{C08F:mmm}}. }\label{F:inf1}
\end{figure}
%%%%%%%%%%%%%%%%%%%%%%%%%%%%%%%%%%%%%%%%%%%%%%%%%%%%%%%%%%%%%%%%%%%%
%%%
By construction, the inflation preserves the maximal number of mutually
crossing and nesting arcs \cite{Reidys:07vac}. Given a $k$-noncrossing tangle,
we can construct a vacillating tableaux, using the following algorithm:
starting from right to left, we take three types of actions: we either
RSK-insert, extract (via Lemma~\ref{L:extract}) or do nothing, depending
on whether we are given an terminus, origin or isolated point of the
inflated tangle, see Figure~\ref{F:psi1}.
%%%%%%%%%%%%%%%%%%%%%%%%%%%%%%%%%%%%%%%%%%%%%%%%%%%%%%%%%%%%%%%%%%%%%%%%
\begin{figure}[ht]
\centerline{\epsfig{file=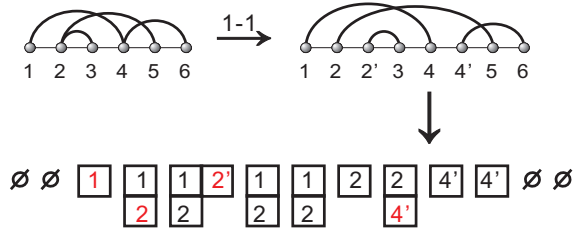,width=0.5\textwidth} \hskip8pt}
\caption{\small From tangled diagrams to vacillating tableaux via
the inflation: for the first tangled diagram in Figure~\ref{F:inf1}
we present its inflation and its unique vacillating
tableaux.}\label{F:psi1}
\end{figure}
%%%
%%%%%%%%%
In fact, the above algorithm has a unique inverse: from a vacillating
tableaux, we can derive a unique tangle, see Figure~{\ref{F:phi1}}.
For $+\square$ steps one simply inserts into the tableaux, does nothing
for $\varnothing$ steps and RSK-extracts (Lemma~\ref{L:extract}) for
$-\square$ steps.
%%
%%%%%%%%%%%%%%%%%%%%%%%%%%%%%%%%%%%%%%%%%%%%%%%%%%%%%%%%%%%%%%%%%%%%%%%%
%%%
%%%%%%%%%%%%%%%%%%%%%%%%%%%%%%%%%%%%%%%%%%%%%%%%%%%%%%%%%%%%%%%%%%%%%%%%
%%%
As result (see Figure~\ref{F:psi1} and Figure~\ref{F:phi1}) we derive the
following theorem \cite{Reidys:07vac}.
%%%
%%%%%%%%%%%%%%%%%%%%%%%%%%%%%%%%%%%%%%%%%%%%%%%%%%%%%%%%%%%%%%%%%%%%%
%%%
\begin{theorem}\label{T:cross}
There exists a bijection between $k$-noncrossing tangled diagrams
and vacillating tableaux of type $\mathcal{V}_{\varnothing}^{2n}$
having shapes $\lambda^i$ with less than $k$ rows.
\end{theorem}
%%%
%%%%%%%%%%%%%%%%%%%%%%%%%%%%%%%%%%%%%%%%%%%%%%%%%%%%%%%%%%%%%%%%%%%%

Theorem~\ref{T:cross} implies bijections between various subclasses
of vacillating tableaux and subclasses of tangles. Most notably the
bijection \cite{Chen:07a} between $k$-noncrossing diagrams and
vacillating tableaux (of empty shape) such that (i)
$\lambda^{0}=\varnothing$ and $\lambda^{2n}=\varnothing,$ and (ii)
$(\lambda^{2i-1},\lambda^{2i})$ is derived from $\lambda^{2i-2}$,
for $1\le i\le n$, by one of the following operations.
$(\varnothing,\varnothing)$: do nothing twice;
$(-\square,\varnothing)$: first remove a square then do nothing;
$(\varnothing,+\square)$: first do nothing then adding a square. We
refer to the latter as $\dagger$-tableaux. Obviously, the latter are
completely determined by the sequence of shapes
$(\lambda^2,\lambda^4,\dots,\lambda^{2n-2})$.

%%%%%%%%%%%%%%%%%%%%%%%%%%%%%%%%%%%%%%%%%%%%%%%%%%%%%%%%%%%%%%%%%%%%%%%%%%%
\subsection{$k$-noncrossing RNA structures}
%%%%%%%%%%%%%%%%%%%%%%%%%%%%%%%%%%%%%%%%%%%%%%%%%%%%%%%%%%%%%%%%%%%%%%%%%%%%
The combinatorics of $k$-noncrossing RNA pseudoknot structures has
been derived in \cite{Reidys:07pseu,Reidys:07asy1}. The set (number)
of $k$-noncrossing, $\sigma$-canonical RNA structures is denoted by
$T_{k,\sigma}(n)$ (${\sf T}_{k,\sigma}^{}(n)$) and let
$f_{k}(n,\ell)$ denote the number of $k$-noncrossing diagrams with
arbitrary arc-length and $\ell$ isolated vertices over $[n]$. It
follows from Theorem~\ref{T:cross}, that the number of
$k$-noncrossing matchings on $[2n]$ equals the number of walks from
$(k-1,k-2,\cdots,1)$ to itself that stay inside the Weyl Chamber
$x_1>x_2>\cdots>x_{k-1}>0$ with steps $\pm e_i$, $1 \le i\le k-1$.
The latter is given by Grabiner {\it et al.}~\cite{Grabiner:93a}. It
is exactly the situation $\eta=\lambda=(k-1,k-2,\cdots,1)$ of
equation $(38)$ in \cite{Grabiner:93a}. As shown in detail in
\cite{Reidys:07pseu}, Lemma~$2$
\begin{eqnarray}
\label{E:ww1} \sum_{n\ge 0} f_{k}(n,0)\cdot\frac{x^{n}}{n!}
&=& \det[I_{i-j}(2x)-I_{i+j}(2x)]|_{i,j=1}^{k-1}\\
\label{E:ww2} \sum_{n\ge
0}\left\{\sum_{\ell=0}^nf_{k}(n,\ell)\right\}\cdot\frac{x^{n}}{n!}
&= & e^{x}\det[I_{i-j}(2x)-I_{i+j}(2x)]|_{i,j=1}^{k-1} ,
\end{eqnarray}
where $I_{r}(2x)=\sum_{j \ge 0}\frac{x^{2j+r}}{{j!(r+j)!}}$ denotes
the hyperbolic Bessel function of the first kind of order $r$.
In particular for $k=2$ and $k=3$ we have
the formulas
\begin{equation}\label{E:2-3}
f_2(n,\ell)  =  \binom{n}{\ell}\,C_{(n-\ell)/2}\quad
\text{\rm and}\quad  f_{3}(n,\ell)=
{n \choose \ell}\left[C_{\frac{n-\ell}{2}+2}C_{\frac{n-\ell}{2}}-
      C_{\frac{n-\ell}{2}+1}^{2}\right].
\end{equation}
In view of $ f_{k}(n,\ell) ={n \choose
\ell} f_{k}(n-\ell,0) $ everything can be reduced to matchings,
where we have the following situation: there exists an
asymptotic approximation of the determinant of hyperbolic Bessel
function for general order $k$ due to \cite{Wang:07} and employing
the subtraction of singularities-principle \cite{Odlyzko:95a} one
can prove \cite{Wang:07}
\begin{equation}\label{E:f-k-imp}
\forall\, k\in\mathbb{N};\qquad  f_{k}(2n,0) \, \sim  \, c_k  \,
n^{-((k-1)^2+(k-1)/2)}\, (2(k-1))^{2n},\qquad \text{\rm where}\ c_k>0 .
\end{equation}
Let ${\bf F}_k(z)=\sum_{n\ge 0}f_k(2n,0)z^{2n}$ denote the generating function
of $k$-noncrossing matchings. Setting
$$
w_0(x)=\frac{x^{2\sigma-2}}{1-x^2+x^{2\sigma}} \quad
\text{\rm and} \quad v_0(x)=1-x+w_0(x)x^2+w_0(x)x^3+w_0(x)x^4
$$
we can now state the following result \cite{Reidys:08ma}.
%%%
%%%%%%%%%%%%%%%%%%%%%%%%%%%%%%%%%%%%%%%%%%%%%%%%%%%%%%%%%%%%%%%%%%%%%%%%%%
%%%
\begin{theorem}\label{T:LLL}
Let $k,\sigma\in \mathbb{N}$, where $k\ge 2,\sigma\ge 3$, let $x$ be
an indeterminate and $\rho_k=\frac{1}{2(k-1)}$ the dominant,
positive real singularity of $\mathbf{F}_k(z)$. Then
$\mathbf{T}_{k,\sigma}^{}(x)$, the generating function of
$k$-noncrossing, $\sigma$-canonical structures, is given by
\begin{equation}
 \mathbf{T}_{k,\sigma}^{}(x)=\frac{1}{v_0(x)}\mathbf{F}_k
\left( \frac{\sqrt{w_0(x)}x}{v_0(x)} \right).
\end{equation}
Furthermore,
\begin{equation}
 \mathbf{T}_{k,\sigma}^{}(n) \sim c_k n^{-(k-1)^2-(k-1)/2}
\left( \frac{1}{\gamma_{k,\sigma}^{}} \right)^n,\ \quad \ \text{\rm
for} \quad k=2,3,4,\ldots,9,
\end{equation}
holds, where $\gamma_{k,\sigma}^{}$  is the minimal positive real
solution of the equation
$\frac{\sqrt{w_0(x)}x}{v_0(x)}=\rho_k=\frac{1}{2(k-1)}$.
\end{theorem}
%%%
%%%%%%%%%%%%%%%%%%%%%%%%%%%%%%%%%%%%%%%%%%%%%%%%%%%%%%%%%%%%%%%%%%%%%%%%%%
%%%

Via Theorem~\ref{T:cross} each $k$-noncrossing, $\sigma$-canonical
structure corresponds to a unique $\dagger$-tableau. We refer to the
set of these tableaux as $\mathfrak{C}$-tableaux.

%%%%%%%%%%%%%%%%%%%%%%%%%%%%%%%%%%%%%%%%%%%%%%%%%%%%%%%%%%%
\subsection{Singularity analysis}
%%%%%%%%%%%%%%%%%%%%%%%%%%%%%%%%%%%%%%%%%%%%%%%%%%%%%%%%%%%%
In view of Theorem~\ref{T:LLL} it is of interest to deduce relations
between the coefficients from the equality of generating functions.
The class of theorems that deal with this deduction are called
transfer-theorems \cite{Flajolet:08}. We use the notation
\begin{equation}\label{E:genau}
\left(f(z)=O\left(g(z)\right) \
\text{\rm as $z\rightarrow \rho$}\right)\quad \Longleftrightarrow \quad
\left(f(z)/g(z) \ \text{\rm is bounded as $z\rightarrow \rho$}\right)
\end{equation}
and if we write $f(z)=O(g(z))$ it is implicitly assumed that $z$
tends to a (unique) singularity. $[z^n]\,f(z)$ denotes the
coefficient of $z^n$ in the power series expansion of $f(z)$ around
$0$.

%%%
%%%%%%%%%%%%%%%%%%%%%%%%%%%%%%%%%%%%%%%%%%%%%%%%%%%%%%%%%%%%%%%%%%%%%%%%%
%%%
\begin{theorem}\label{T:transfer1}{\bf }
\cite{Flajolet:08} Let $f(z),g(z)$ be $D$-finite functions with
unique dominant singularity $\rho$ and suppose $f(z) = O( g(z))$ for
$z\rightarrow \rho$. Then we have
\begin{equation}
[z^n]f(z)= \,K \,\left(1-O\left(\frac{1}{n}\right)\right)\,
[z^n]g(z) ,
\end{equation}
where $K$ is some constant.
\end{theorem}
%%%
%%%%%%%%%%%%%%%%%%%%%%%%%%%%%%%%%%%%%%%%%%%%%%%%%%%%%%%%%%%%%%%%%%%%%%%%%
%%%
Theorem~\ref{T:transfer1} and eq.~(\ref{E:f-k-imp}) imply
\begin{eqnarray}
{\bf F}_k(z)=
\begin{cases}
O((1-\frac{z}{\rho_k})^{(k-1)^2+(k-1)/2-1} \ln(1-\frac{z}{\rho_k}))
& \text{\rm for $k$ odd,
$z\rightarrow \rho_k$} \\
O((1-\frac{z}{\rho_k})^{(k-1)^2+(k-1)/2-1})             & \text{\rm
for $k$
 even,
$z\rightarrow \rho_k$,}
\end{cases}\label{E:F_k}
\end{eqnarray}
in accordance with basic structure theorems for singular expansions
of $D$-finite functions \cite{Flajolet:08}. Furthermore,
Theorem~\ref{T:transfer1}, eq.~(\ref{E:f-k-imp}) and the so called
subcritical case of singularity analysis \cite{Flajolet:08}, VI.9.,
p.~411, imply the following result tailored for our functional
equations \cite{Reidys:07lego}. Let $\rho_k$ denote the dominant
positive real singularity of ${\bf F}_{k}(z)$.
%%%
%%%%%%%%%%%%%%%%%%%%%%%%%%%%%%%%%%%%%%%%%%%%%%%%%%%%%%%%%%%%%%%%%%%%%%%%%
%%%
\begin{theorem}\label{T:realdeal}
Suppose $\vartheta_{\sigma}(z)$ is algebraic over $K(z)$, analytic
for $\vert z\vert <\delta$ and satisfies $\vartheta_{\sigma}(0)=0$.
Suppose further $\gamma_{k,\sigma}$ is the real unique solution with
minimal modulus $<\delta$ of the two equations
$\vartheta_{\sigma}(z)=\rho_k$ and $\vartheta_{\sigma}(z)=-\rho_k$.
Then
\begin{equation}
[z^n]\,{\bf F}_k(\vartheta_{\sigma}(z)) \sim c_k  \,
n^{-((k-1)^2+(k-1)/2)}\, \left(\gamma_{k,\sigma}^{-1}\right)^n .
\end{equation}
\end{theorem}
%%%
%%%%%%%%%%%%%%%%%%%%%%%%%%%%%%%%%%%%%%%%%%%%%%%%%%%%%%%%%%%%%%%%%%%%%%%%%
%%%

The below continuity theorem of discrete limit laws will be used in
the proofs of Theorem~\ref{T:centra2} and Theorem~\ref{T:central}.
It ensures that under certain conditions the point-wise convergence
of probability generating functions implicates the convergence of
its coefficients.
%%%
%%%%%%%%%%%%%%%%%%%%%%%%%%%%%%%%%%%%%%%%%%%%%%%%%%%%%%%%%%%%%%%%%%%%%%%%%
%%%
\begin{theorem}{\label{T:continuity}}
Let $u$ be an indeterminate and $\Omega$ be a set contained in the
unit disc, having at least one accumulation point in the interior of
the disc. Assume $P_n(u)=\sum_{k\ge 0}p_{n,k}u^k$ and
$q(u)=\sum_{k\ge 0}q_k u^k$ such that $\lim_{n\rightarrow
\infty}P_n(u)=q(u)$ for each $u\in\Omega$ holds. Then we have for
any finite $k$,
\begin{equation}
\lim_{n\rightarrow\infty}p_{n,k}=q_k \quad \ \text{\it and }\quad \
\lim_{n\rightarrow \infty}\sum_{j\le k}p_{n,j}=\sum_{j\le k}q_j.
\end{equation}
\end{theorem}
%%%
%%%%%%%%%%%%%%%%%%%%%%%%%%%%%%%%%%%%%%%%%%%%%%%%%%%%%%%%%%%%%%%%%%%%%%%%%
%%%
%%%
%%%%%%%%%%%%%%%%%%%%%%%%%%%%%%%%%%%%%%%%%%%%%%%%%%%%%%%%%%%%%%%%%%%%%%
%%%
\section{Irreducible substructures}\label{S:gap}

%%%
%%%%%%%%%%%%%%%%%%%%%%%%%%%%%%%%%%%%%%%%%%%%%%%%%%%%%%%%%%%%%%%%%%%%%%%%%%%%%%
%%%
In the following we shall identify a $\mathfrak{C}$-tableaux with
the subsequence of even-indexed shapes, i.e.~the sequence
$(\lambda^2, \dots,\lambda^{2n-2})$. Subsequences of two or more
consecutive $\varnothing$-shapes result from the elementary move
$(\varnothing, \varnothing)$. For instance, consider the
$\mathfrak{C}$-tableaux
\begin{center}
\setlength{\unitlength}{3pt}
\begin{picture}(110,5)
\put(0,0){$\varnothing$} \put(0,6.0){$\lambda^0$}
\put(3.5,1.0){\vector(1,0){10}}
\put(3.5,2.0){$(\varnothing,\varnothing)$} \put(15,0){$\varnothing$}
\put(15,6.0){$\lambda^2$}\put(18.5,1.0){\vector(1,0){12}}
\put(18.5,2.0){$(\varnothing,+\Box)$}\put(32,-0.5){\framebox(4,4)}
\put(37,1.0){\vector(1,0){12}}\put(37,2.0){$(\varnothing,+\Box)$}
\put(33,6.0){$\lambda^4$} \put(50.5,-3){\framebox(4,8)}
\put(50.5,1){\line(1,0){4}}\put(52,6){$\lambda^6$}
\put(55.5,1.0){\vector(1,0){12}}
\put(56,2.0){$(\varnothing,\varnothing)$}
\put(69.5,-3){\framebox(4,8)}\put(69.5,1){\line(1,0){4}}
\put(71,6){$\lambda^8$}\put(74.5,1.0){\vector(1,0){12}}
\put(75,2.0){$(-\Box,\varnothing)$} \put(88.5,-0.5){\framebox(4,4)}
\put(90,6){$\lambda^{10}$}\put(93.5,1){\vector(1,0){12}}
\put(108,6){$\lambda^{12}$} \put(94,2.0){$(-\Box,\varnothing)$}
\put(108,-0.5){$\varnothing$}
\end{picture}
\end{center}
The above tableaux splits at $\lambda^2=\varnothing$ into two
$\mathfrak{C}$-subtableaux, i.e.~
\begin{center}
\setlength{\unitlength}{3pt}
\begin{picture}(120,5)
\put(0,0){$\varnothing$} \put(0,6.0){$\lambda^0$}
\put(3.5,1.0){\vector(1,0){10}}
\put(3.5,2.0){$(\varnothing,\varnothing)$} \put(15,0){$\varnothing$}
\put(15,6.0){$\lambda^2$} \put(20,3){and} \put(27,0){$\varnothing$}
\put(27,6.0){$\lambda^2$}\put(30.5,1.0){\vector(1,0){12}}
\put(30.5,2.0){$(\varnothing,+\Box)$}\put(44,-0.5){\framebox(4,4)}
\put(49,1.0){\vector(1,0){12}}\put(49,2.0){$(\varnothing,+\Box)$}
\put(45,6.0){$\lambda^4$} \put(62.5,-3){\framebox(4,8)}
\put(62.5,1){\line(1,0){4}}\put(64,6){$\lambda^6$}
\put(67.5,1.0){\vector(1,0){12}}
\put(68,2.0){$(\varnothing,\varnothing)$}
\put(81.5,-3){\framebox(4,8)}\put(81.5,1){\line(1,0){4}}
\put(83,6){$\lambda^8$}\put(86.5,1.0){\vector(1,0){12}}
\put(87,2.0){$(-\Box,\varnothing)$} \put(100.5,-0.5){\framebox(4,4)}
\put(102,6){$\lambda^{10}$}\put(105.5,1){\vector(1,0){12}}
\put(120,6){$\lambda^{12}$} \put(106,2.0){$(-\Box,\varnothing)$}
\put(120,-0.5){$\varnothing$}
\end{picture}
\end{center}
We call a sequence of consecutive $\varnothing$-shapes of length
$(r+1)$, $(\varnothing,\dots,\varnothing)$ a gap of length $r$.
Theorem~\ref{T:cross} implies that these $\varnothing$-gaps
correspond uniquely to the gaps of diagrams, introduced in
Section~\ref{S:back}. A $*$-tableaux is a $\mathfrak{C}$-tableaux,
with the property $\lambda^i\ne \varnothing$ for $2 \le i\le 2n-2$.
It is evident that a $*$-tableaux corresponds via the bijection of
Theorem~\ref{T:cross} to an irreducible $k$-noncrossing,
$\sigma$-canonical RNA structure. For instance,
\begin{figure}[ht]
\centerline{\epsfig{file=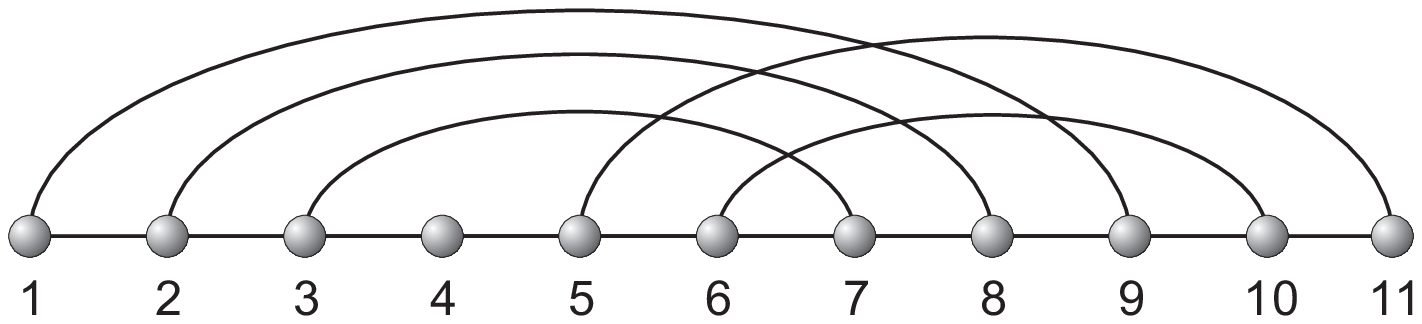,width=0.5\textwidth}
\hskip8pt}\label{F:pb}
\begin{center}
\setlength{\unitlength}{2.5pt}
\begin{picture}(200,5)
\put(0,-0.5){$\varnothing$}
\put(3.5,1.0){\vector(1,0){5}}\put(0,6){$\lambda^0$}
\put(9,-0.5){\framebox(4,4)}\put(10,6){$\lambda^2$}
\put(14,1.0){\vector(1,0){5}}
\put(19.5,-0.5){\framebox(8,4)}\put(23.5,-0.5){\line(0,1){4}}
\put(22,6){$\lambda^4$} \put(28.5,1.0){\vector(1,0){5}}
\put(34,-0.5){\framebox(12,4)}\put(38,-0.5){\line(0,1){4}}
\put(42,-0.5){\line(0,1){4}}\put(40,6){$\lambda^6$}
\put(47,1.0){\vector(1,0){5}}\put(52.5,-0.5){\framebox(12,4)}
\put(56.5,-0.5){\line(0,1){4}} \put(60.5,-0.5){\line(0,1){4}}
\put(58.5,6){$\lambda^8$}\put(65.5,1.0){\vector(1,0){5}}
\put(71,-0.5){\framebox(12,4)}\put(75,-4.5){\line(0,1){8}}
\put(79,-0.5){\line(0,1){4}}\put(70.9,-4.5){\line(0,1){4}}
\put(71,-4.5){\line(1,0){4}}\put(75,6){$\lambda^{10}$}
\put(84,1.0){\vector(1,0){5}}
\put(89.5,-0.5){\framebox(12,4)}\put(93.5,-4.5){\line(0,1){8}}
\put(97.5,-4.5){\line(0,1){8}}\put(89.4,-4.5){\line(0,1){4}}
\put(89.5,-4.5){\line(1,0){8}}\put(93.5,6){$\lambda^{12}$}
\put(102.5,1.0){\vector(1,0){5}} \put(108.5,-4.5){\framebox(8,8)}
\put(108.5,-0.5){\line(1,0){8}}\put(112.5,-4.5){\line(0,1){8}}
\put(111.5,6){$\lambda^{14}$}\put(117.5,1.0){\vector(1,0){5}}
\put(123,-0.5){\framebox(8,4)}\put(127,-4.5){\line(0,1){8}}
\put(122.9,-4.5){\line(0,1){4}}\put(123,-4.5){\line(1,0){4}}
\put(126,6){$\lambda^{16}$} \put(133.5,1.0){\vector(1,0){5}}
\put(139.5,-0.5){\framebox(8,4)}\put(143.5,-0.5){\line(0,1){4}}
\put(141.5,6){$\lambda^{18}$} \put(148.5,1.0){\vector(1,0){5}}
\put(154.5,-0.5){\framebox(4,4)}\put(154,6){$\lambda^{20}$}
\put(161,1.0){\vector(1,0){5}} \put(166.5,-0.5){$\varnothing$}
\put(166.5,6){$\lambda^{22}$}
\end{picture}
\end{center}
\end{figure}

Obviously, any $\mathfrak{C}$-tableaux can be uniquely decomposed
into a sequences of gaps and $*$-tableaux. For instance,
\begin{center}
\setlength{\unitlength}{3pt}
\begin{picture}(100,5)
\put(-8,0){$\varnothing$}\put(-8,6){$\lambda^0$}
\put(-5.5,1.0){\vector(1,0){5}} \put(0,0){$\varnothing$}
\put(0,6){$\lambda^2$} \put(3.5,1.0){\vector(1,0){5}}
\put(9,-0.5){\framebox(4,4)}\put(10,6){$\lambda^4$}
\put(14,1.0){\vector(1,0){5}}
\put(19.5,-0.5){\framebox(8,4)}\put(23.5,-0.5){\line(0,1){4}}
\put(22,6){$\lambda^6$} \put(28.5,1.0){\vector(1,0){5}}
\put(34.5,-0.5){\framebox(4,4)} \put(35.5,6){$\lambda^8$}
\put(39.5,1.0){\vector(1,0){5}}\put(45.5,-3){\framebox(4,8)}
\put(45.5,1){\line(1,0){4}}\put(46.5,6){$\lambda^{10}$}
\put(50.5,1.0){\vector(1,0){5}}
\put(56.5,-0.5){\framebox(4,4)}\put(57.5,6){$\lambda^{12}$}
\put(61.5,1.0){\vector(1,0){5}}
\put(67.5,0){$\varnothing$}\put(68.5,6){$\lambda^{14}$}
\put(71.5,1.0){\vector(1,0){5}}
\put(77.5,0){$\varnothing$}\put(77.5,6){$\lambda^{16}$}
\put(81,1.0){\vector(1,0){5}}
\put(87.5,0){$\varnothing$}\put(87.5,6){$\lambda^{18}$}
\put(92.5,1.0){\vector(1,0){5}} \put(98.5,0){$\varnothing$}
\put(98.5,6){$\lambda^{20}$}
\end{picture}
\end{center}
splits into the gap $(0,2)$, the $*$-tableaux over $(2,14)$ and the
gap $(14,20)$. Let $\delta_{n,j}^{(k)}$ denote the number of
$\mathfrak{C}$-tableaux of length $2n$ with less than $k$ rows, containing
exactly $j$ $*$-tableaux. Furthermore, let
\begin{equation}\label{E:JJ}
{\bf U}_{k}(z,u)=\sum_{n\ge 0}\sum_{j\ge 0}\delta_{n,j}^{(k)}u^jz^n,
\end{equation}
and $\delta_{n}^{(k)}=\sum_{j\ge 0}\delta_{n,j}^{(k)}$. We set
${\bf T}_{k}(z)={\bf T}_{k,\sigma}(z)=\sum_{n\ge 0}\delta_{n}^{(k)}z^n$
and denote the generating function of $*$-tableaux by ${\bf R}_k(z)$.

%%%
%%%%%%%%%%%%%%%%%%%%%%%%%%%%%%%%%%%%%%%%%%%%%%%%%%%%%%%%%%%%%%%%%%%%%%%%%%%%%
%%%
\begin{lemma}\label{L:bistar}
The bivariate generating function of the number of $\mathfrak{C}$-tableaux
of length $2n$ with less than $k$ rows, which contain exactly $i$ $*$-tableaux,
is given by
\begin{eqnarray*}
{\bf U}_{k}(z,u) &=&\frac{\frac{1}{1-z}}
{1-u\left(1-\frac{1}{(1-z){\bf T}_k(z)}\right)}.\\
\end{eqnarray*}
\end{lemma}
%%%
%%%%%%%%%%%%%%%%%%%%%%%%%%%%%%%%%%%%%%%%%%%%%%%%%%%%%%%%%%%%%%%%%%%%%%%%%%%%%
%%%
\begin{proof}
Since each $\mathfrak{C}$-tableau can be uniquely decomposed into a
sequence of gaps and $*$-tableaux we obtain for fixed $j$
\begin{eqnarray}
\sum_{n\ge j}\delta_{n,j}z^n &=& {\bf
R}_k(z)^j\left(\frac{1}{1-z}\right)^{j+1}.
\end{eqnarray}
As a result the bivariate generating function of $\delta_{n,j}$ is given by
\begin{equation}
{\bf U}_k(z,u)=\sum_{j\ge 0}\sum_{n\ge j}\delta_{n,j}z^n u^j =
\sum_{j\ge 0}{\bf
R}_k(z)^j\left(\frac{1}{1-z}\right)^{j+1}u^j\\
=\frac{1}{1-z-u{\bf R}_k(z)}.
\end{equation}
Setting $u=1$ we derive
\begin{eqnarray}
{\bf T}_k(z)={\bf U}_k(z,1)&=&\frac{1}{1-z-{\bf R}_k(z)}
\end{eqnarray}
which allows us to express the generating function of $*$-tableaux via
${\bf T}_k(z)$
\begin{eqnarray}\label{E:well}
{\bf R}_k(z)=1-z-\frac{1}{{\bf T}_k(z)}.
\end{eqnarray}
Consequently, ${\bf U}_k(z,u)$ is given by
\begin{equation}
{\bf U}_k(z,u) = \frac{1}{1-z-u{\bf R}_k(z)}
= \frac{\frac{1}{1-z}}{1-u\left(1-\frac{1}{(1-z){\bf
T}_k(z)}\right)}
\end{equation}
and the lemma follows.
\end{proof}
%%%
%%%%%%%%%%%%%%%%%%%%%%%%%%%%%%%%%%%%%%%%%%%%%%%%%%%%%%%%%%%%%%%%%%%%%%%%%%%%%
%%%
Setting $g(z)=\frac{1}{1-z}$ and $h(z)=1-\frac{1}{(1-z){\bf T}_k(z)}$,
Lemma~\ref{L:bistar} implies
\begin{eqnarray}\label{E:express}
{\bf U}_{k}(z,u) &=&g(z)\cdot\frac{1}{1-uh(z)}=g(z)\cdot g(uh(z)).
\end{eqnarray}
Let $\xi_n^{(k)}$ be a r.v.~such that $\mathbb{P}(\xi_{n}^{(k)}=i) =
\frac{\delta_{n,i}^{(k)}}{\delta_{n}^{(k)}}$ and let $\rho_{p}$ and
$\rho_{w}$ denote the radius of convergence of the power series
$p(z)$ and $w(z)$, respectively. We denote
$\tau_{w}=\lim_{z\rightarrow\rho_{w}^{-}}w(z)$ and call a function
$F(z,u)=p(u\cdot w(z))$ subcritical if and only if
$\tau_{w}<\rho_{p}$.

%%%
%%%%%%%%%%%%%%%%%%%%%%%%%%%%%%%%%%%%%%%%%%%%%%%%%%%%%%%%%%%%%%%%%%%%%%%%%%
%%%
\begin{theorem}{\label{T:centra2}}
Let $\alpha_k$ be the real positive dominant singularity of ${\bf
T}_{k}(z)$ and $\tau_k=1-\frac{1}{(1-\alpha_k){\bf
T}_{k}(\alpha_k)}$. Then the r.v.~$\xi_n^{(k)}$ satisfies the
discrete limit law
\begin{equation}
\lim_{n\rightarrow
\infty}\mathbb{P}(\xi_{n}^{(k)}=i)=q_i {\mbox \quad where
\quad} q_i=\frac{(1-\tau_k)^2}{\tau_k}\,i \tau_k^{i}.
\end{equation}
That is, $\xi_n^{(k)}$ is determined by the density function of a
$\Gamma(-\ln\tau_k,2)$-distribution.
Furthermore, the probability generating function of the limit distribution
$q(u)=\sum_{n\ge 1}q_i u^i$ satisfies $q(u)=\frac{u(1-\tau_k)^2}
{(1-\tau_k u)^2}$.
\end{theorem}
%%%
%%%%%%%%%%%%%%%%%%%%%%%%%%%%%%%%%%%%%%%%%%%%%%%%%%%%%%%%%%%%%%%%%%%%%%%%%
%%%
\begin{proof}
Since $g(z)=\frac{1}{1-z}$ and $h(z)=1-\frac{1}{(1-z){\bf T}_k(z)}$
have non negative coefficients and $h(0)=0$, the composition
$g(h(z))$ is well defined as formal power series. According to
eq.~(\ref{E:express}) we may express ${\bf U}_k(z,u)$ as $ {\bf
U}_k(z,u) = g(z)g(uh(z))$. For $z=\alpha_k$ we have
$\tau_k=1-\frac{1}{(1-\alpha_k){\bf T}_k(\alpha_k)}<1=\rho_{g}$,
i.e.~we are given the subcritical case.\\
%%%%%%%%%%%%%%%%%%%%%%%%%%%%%%%%%%%%%%%%%%%%%%%%%%%%%%%%%%%%%%%%%%%%%%%%%%%%%%
{\it Claim $1$.} $h(z)$ has a singular expansion at its dominant
singularity $z=\alpha_k$ and there exists some constant $c_k>0$ such
that
\begin{equation}
h(z)=
\begin{cases}
\tau_k-c_k\left(1-\frac{z}{\alpha_k}\right)^{\mu}
\ln\left(1-\frac{z}{\alpha_k}\right)(1+o(1)) &
\text{\rm for}\ k\equiv 1 \mod 2\\
\tau_k-c_k\left(1-\frac{z}{\alpha_k}\right)^{\mu}(1+o(1)) &
\text{\rm for}\ k\equiv 0 \mod 2
\end{cases}
\end{equation}
for $z\rightarrow \alpha_k$ and $\mu=(k-1)^2+\frac{k-1}{2}-1$.\\
%%%%%%%%%%%%%%%%%%%%%%%%%%%%%%%%%%%%%%%%%%%%%%%%%%%%%%%%%%%%%%%%%%%%%%%%%%%%%
Since ${\bf F}_k(z)$ is $D$-finite, the composition ${\bf
F}_k(\vartheta(z))$ where $\vartheta(z)=
\frac{\sqrt{w_0(z)}z}{v_0(z)}$ and $\vartheta(0)=0$, is also
$D$-finite \cite{Stanley:80}. As a result, ${\bf T}_k(z)$ is, being
a product of the two $D$-finite functions $\frac{1}{v_0(z)}$ and
${\bf F}_k(\vartheta(z))$, $D$-finite. We view of $\frac{1}{{\bf
T}_k(z)}$ is the composition of the outer function ${\bf
H}(z)=\frac{1}{1+z}$ and inner function ${\bf T}_k(z)-1$, where
${\bf T}_k(0)-1=0$. We conclude from this, that
$h(z)=1-\frac{1}{(1-z){\bf T}_k(z)}$ is $D$-finite. $h(z)$ is
analytic at $z=0$ and its $D$-finiteness guarantees that $h(z)$ has
an analytic continuation in some simply connected
$\Delta_{\alpha_k}$-domain containing zero \cite{Stanley:80}.
Consequently, the singular expansion of $h(z)$ at $z=\alpha_k$ does
exist and
\begin{eqnarray*}
h(z)&=&\tau_k+h'(\alpha_k)(z-\alpha_k)
+\frac{h''(\alpha_k)}{2!}(z-\alpha_k)^2+\cdots\\
&=&\tau_k+\frac{{\bf T}_{k}'(\alpha_k)}{{\bf T}_{k}^2(\alpha_k)}
(z-\alpha_k)+\left[\frac{{\bf T}_{k}''(\alpha_k)}{2{\bf
T}_{k}^2(\alpha_k)}-\frac{({\bf T}_{k}'(\alpha_k))^2}{{\bf
T}_{k}^3(\alpha_k)}\right](z-\alpha_k)^2+\cdots
\end{eqnarray*}
We next observe that Theorem~\ref{T:transfer1}, the singular
expansion of ${\bf F}_k(z)$ at $\rho_k$ and Theorem~\ref{T:realdeal}
imply
\begin{eqnarray}\label{E:asy}
{\bf T}_k(z)=
\begin{cases}
O((1-\frac{z}{\alpha_k})^{(k-1)^2+(k-1)/2-1}
\ln(1-\frac{z}{\alpha_k})) & \text{\rm for $k$ odd,
$z\rightarrow \alpha_k$} \\
O((1-\frac{z}{\alpha_k})^{(k-1)^2+(k-1)/2-1})             &
\text{\rm for $k$
 even, $z\rightarrow \alpha_k$.}
\end{cases}\label{E:W_k}
\end{eqnarray}
Suppose first that $k\equiv 1\mod 2$ and set
$\mu=(k-1)^2+\frac{k-1}{2}-1$. For $z\rightarrow\alpha_k$,
eq.~(\ref{E:asy}) guarantees
\begin{eqnarray}\label{E:g1}
{\bf T}_k(z)=\ell(\alpha_k)\left(1-\frac{z}{\alpha_k}\right)^{\mu}
\ln\left(1-\frac{z}{\alpha_k}\right)+r(\alpha_k),
\end{eqnarray}
where $\ell(\alpha_k)<0$. Since $\ell(\alpha_k)<0$, ${\bf T}_k(z)$
is a power series with positive coefficients and in view of $\mu\ge
\frac{1}{2}$, for any $k\ge 2$, ${\bf T}_k(\alpha_k)<\infty$.
Accordingly, we obtain for $z\rightarrow\alpha_k$
\begin{align}\label{E:g2}
{\bf T}'_k(z)&=-\frac{\mu}{\alpha_k}\cdot\ell(\alpha_k)
\left(1-\frac{z}{\alpha_k}\right)^{\mu-1}
\ln\left(1-\frac{z}{\alpha_k}\right)-
\frac{\ell(\alpha_k)}{\alpha_k}\left(1-\frac{z}{\alpha_k}\right)^{\mu-1}\\
\label{E:g3}{\bf
T}''_k(z)&=\frac{\mu(\mu-1)}{\alpha_k^2}\cdot\ell(\alpha_k)
\left(1-\frac{z}{\alpha_k}\right)^{\mu-2}
\ln\left(1-\frac{z}{\alpha_k}\right)+\frac{(2\mu-1)
\ell(\alpha_k)}{\alpha_k^2}\left(1-\frac{z}{\alpha_k}\right)^{\mu-2}.
\end{align}
Eq.~(\ref{E:g2}) and eq.~(\ref{E:g3}) imply
\begin{eqnarray*}
h'(\alpha_k)(z-\alpha_k) &=&\frac{\mu\ell(\alpha_k)}{{\bf
T}_k^2(\alpha_k)}\left(1-\frac{z}{\alpha_k}\right)^{\mu}
\ln\left(1-\frac{z}{\alpha_k}\right)(1+o(1))\\
\frac{h''(\alpha_k)}{2}(z-\alpha_k)^2 &=&\frac{
\mu(\mu-1)\ell(\alpha_k)}{2{\bf
T}_k^2(\alpha_k)}\left(1-\frac{z}{\alpha_k}\right)^{\mu}
\ln\left(1-\frac{z}{\alpha_k}\right)(1+o(1))\\
\frac{h'''(\alpha_k)}{3!}(z-\alpha_k)^3&=& \frac{
\mu(\mu-1)(\mu-2)\ell(\alpha_k)}{3!{\bf
T}_k^2(\alpha_k)}\left(1-\frac{z}{\alpha_k}\right)^{\mu}
\ln\left(1-\frac{z}{\alpha_k}\right)(1+o(1)).
\end{eqnarray*}
We proceed by computing
\begin{eqnarray*}
h(z)&=& \tau_k+\frac{{\bf T}_{k}'(\alpha_k)}{{\bf
T}_{k}^2(\alpha_k)} (z-\alpha_k)+\left[\frac{{\bf
T}_{k}''(\alpha_k)}{2{\bf T}_{k}^2(\alpha_k)}-\frac{({\bf
T}_{k}'(\alpha_k))^2}{{\bf
T}_{k}^3(\alpha_k)}\right](z-\alpha_k)^2+\cdots\\
&=&\tau_k+\frac{\ell(\alpha_k)}{{\bf
T}_k^2(\alpha_k)}\left[\mu+\frac{\mu(\mu-1)}{2}
+\frac{\mu(\mu-1)(\mu-2)}{3!}+\cdots\right]
\left(1-\frac{z}{\alpha_k}\right)^{\mu}
\ln\left(1-\frac{z}{\alpha_k}\right)(1+o(1))\\
&=&\tau_k-c_k\left(1-\frac{z}{\alpha_k}\right)^{\mu}
\ln\left(1-\frac{z}{\alpha_k}\right)(1+o(1)), \quad \text{\rm where}
\ c_k >0.
\end{eqnarray*}
The case $k\equiv 0\mod 2$ is proved analogously and Claim $1$
follows. ${\bf U}_k(z,1)$ is as the product of $g(z)$ and $g(h(z))$
where $h(0)=0$, $D$-finite and has a singular expansion at
$z=\alpha_k$. Without loss of generality, we may restrict ourselves
in the following to the case $k\equiv 1\mod 2$ and proceed by
computing
\begin{eqnarray*}
{\bf U}_{k}(z,1)&=&g(\alpha_k)g(\tau_k)+(g\cdot
g(h))'(\alpha_k)(z-\alpha_k) +\frac{(g\cdot
g(h))''(\alpha_k)}{2!}(z-\alpha_k)^2+\cdots\\
&=& g(\alpha_k)g(\tau_k)-c_kg(\alpha_k)g'(\tau_k)
\left(1-\frac{z}{\alpha_k}\right)^{\mu}
\ln\left(1-\frac{z}{\alpha_k}\right)(1+o(1)).
\end{eqnarray*}
Therefore we derive
\begin{equation}
[z^n]{\bf U}_{k}(z,1)=-c_kg(\alpha_k)g'(\tau_k)\alpha_k^{-n}
n^{-\mu-1}(1+o(1)).
\end{equation}
For any fixed $u\in(0,1)$ the singular expansion of ${\bf
U}_{k}(z,u)$ at $z=\alpha_k$ is given by
\begin{align*}
{\bf U}_{k}(z,u)&=g(\alpha_k)g(u\tau_k)+(g\cdot
g(uh))'(\alpha_k)(z-\alpha_k) +\frac{(g\cdot
g(uh))''(\alpha_k)}{2!}(z-\alpha_k)^2+\cdots\\
&=g(\alpha_k)g(u\tau_k)-c_kug(\alpha_k)
g'(u\tau_k)\left(1-\frac{z}{\alpha_k}\right)^{\mu}
\ln\left(1-\frac{z}{\alpha_k}\right)(1+o(1))
\end{align*}
and we consequently obtain, setting
$\tau_k=1-\frac{1}{(1-\alpha_k){\bf T}_{k}(\alpha_k)}$
\begin{equation}\label{E:guck}
\lim_{n\rightarrow \infty}\frac{[z^n]{\bf U}_{k}(z,u)}{[z^n]{\bf
U}_{k}(z,1)} =\frac{ug'(u\tau_k)}{g'(\tau_k)}=
\frac{u(1-\tau_k)^2}{(1-\tau_k u)^2}=q(u).
\end{equation}
In view of eq.~(\ref{E:guck}) and
$[u^i]q(u)=\frac{(1-\tau_k)^2}{\tau_k}\, i \tau_k^{i}=q_i$,
Theorem~\ref{T:continuity} implies the
discrete limit law
\begin{equation}\label{E:WW}
\lim_{n\rightarrow \infty}\mathbb{P}(\xi_{n}^{(k)}=i)=
\lim_{n\rightarrow\infty}\frac{\delta_{n,i}^{(k)}}
{\delta_{n}^{(k)}}=q_i \qquad \text{\rm where} \quad
q_i=\frac{(1-\tau_k)^2}{\tau_k}\, i \tau_k^{i}.
\end{equation}
Since the density function of a $\Gamma(\lambda,r)$-distribution is
given by
\begin{equation}\label{E:gamma}
f_{\lambda,r}(x)=\left\{
\begin{array}{ll}
\frac{\lambda^r}{\Gamma(r)}x^{r-1}e^{-\lambda x}, & x>0\\
 0 & x\le 0,
\end{array}
\right.
\end{equation}
where $\lambda>0$ and $r>0$, we obtain, setting $r=2$ and $\lambda=-\ln
\tau_k>0$
\begin{eqnarray*}
\lim_{n\to \infty}
\mathbb{P}(\xi_{n}^{(k)}=i) &=& \frac{(1-\tau_k)^2}{\tau_k}(i\cdot \tau_k^i)
\; = \;\frac{(1-\tau_k)^2}{\tau_k}\frac{1}{(\ln \tau_k)^2}(\ln
\tau_k)^2i\cdot \tau_k^i\\
& = & \frac{(1-\tau_k)^2}{\tau_k}\frac{1}{(\ln \tau_k)^2}\ f_{-\ln\tau_k,2}(i)
\end{eqnarray*}
and the proof of the theorem is complete.
\end{proof}
%%%%
%%%%%%%%%%%%%%%%%%%%%%%%%%%%%%%%%%%%%%%%%%%%%%%%%%%%%%%%%%%%%%%%%%%%
%%%%

\section{The limit distribution of nontrivial returns}\label{S:empty}

%%%
%%%%%%%%%%%%%%%%%%%%%%%%%%%%%%%%%%%%%%%%%%%%%%%%%%%%%%%%%%%%%%%%%%%%%%
%%%
Let $\beta_{n}^{(k)}$ denote the number of $\mathfrak{C}$-tableaux
of length $2n$, which are in correspondence to $k$-noncrossing,
$\sigma$-canonical RNA structures. Let $\beta_{n,i}^{(k)}$ denote
the number of $\mathfrak{C}$-tableaux of length $2n$, having
exactly $i$ $\varnothing$-shapes contained in the sequence
$(\lambda^2,\dots,\lambda^{2n})$.
Let ${\bf W}_k(z,u)$ denote the bivariate generating function of
$\beta_{n,i}^{(k)}$. Then $\beta_{n,j}^{(k)}=[z^nu^j]{\bf W}_{k}(z,u)$
and ${\bf W}_{k}(z,u)=\sum_{j\ge 0}\sum_{n\ge j}\beta_{n,j}z^n u^j$.
Furthermore we set $\beta_{n}^{(k)}=[z^n]{\bf W}_k(z,1)$.
%%%
%%%%%%%%%%%%%%%%%%%%%%%%%%%%%%%%%%%%%%%%%%%%%%%%%%%%%%%%%%%%%%%%%%%%%%%%%%
%%%
\begin{lemma}{\label{L:bivariate}}
The bivariate generating function of the number of $\mathfrak{C}$-tableaux
of length $2n$, with less than $k$ rows, containing exactly $i$
$\varnothing$-shapes, is given by
\begin{eqnarray}
{\bf W}_k(z,u)&=&\frac{1}{1-u\left(1-\frac{1}{{\bf T}_k(z)}\right)}.
\end{eqnarray}
\end{lemma}
%%%
%%%%%%%%%%%%%%%%%%%%%%%%%%%%%%%%%%%%%%%%%%%%%%%%%%%%%%%%%%%%%%%%%%%%%%%%%%
%%%
\begin{proof}
Suppose the $\mathfrak{C}$-tableaux $(\lambda^2,\dots,\lambda^{2n})$
contains exactly $i$ $\varnothing$-shapes. These $\varnothing$-shapes split
$(\lambda^2,\dots,\lambda^{2n})$ uniquely into exactly $i$
$\mathfrak{C}$-subtableaux, each of which either being a gap
of length $2$ or an irreducible $*$-tableaux. We conclude from this, that for
fixed $j$
\begin{eqnarray}
\sum_{n\ge j}\beta_{n,j}z^n &=& \left(z+{\bf R}_{k}(z)\right)^j
\end{eqnarray}
holds. Therefore the bivariate generating function ${\bf W}_k(z,u)$ satisfies
\begin{eqnarray*}
{\bf W}_k(z,u)=\sum_{j\ge 0}\sum_{n\ge j}\beta_{n,j}z^n u^j
&=&\sum_{j\ge 0} \left(z+{\bf R}_{k}(z)\right)^j u^j\\
&=&\frac{1}{1-u(z+{\bf R}_k(z))}\\
&=&\frac{1}{1-u(1-\frac{1}{{\bf T}_k(z)})},
\end{eqnarray*}
where the last equality follows from eq.~(\ref{E:well}), proving the lemma.
\end{proof}
We set $g(z)=\frac{1}{1-z}$, $h(z)=1-\frac{1}{{\bf T}_k(z)}$ and
let $\eta_{n}^{(k)}$ denote the random variable having probability
distribution
$
\mathbb{P}(\eta_{n}^{(k)}=i)=\frac{\beta_{n,i}^{(k)}}{\beta_{n}^{(k)}}
$.
In case of ${\bf W}_{k}(z,u)=g(uh(z))$ we have $\rho_{g}=1$ while
$\tau_{h}<1$, i.e.~we are given the subcritical case. In our next
theorem, we prove that the limit distribution of $\eta_{n}^{(k)}$ is
determined by the density function of a $\Gamma(\lambda,r)$-distribution.
%%%
%%%%%%%%%%%%%%%%%%%%%%%%%%%%%%%%%%%%%%%%%%%%%%%%%%%%%%%%%%%%%%%%%%%%%%%%%
%%%
\begin{theorem}{\label{T:central}}
Let $\alpha_k$ denote the real, positive, dominant singularity of
${\bf T}_{k}(z)$ and let $\tau_k=1-\frac{1}{{\bf T}_{k}(\alpha_k)}$.
Then the r.v.~$\eta_{n}^{(k)}$ satisfies the discrete
limit law
\begin{equation}\label{E:one}
\lim_{n\rightarrow
\infty}\mathbb{P}(\eta_{n}^{(k)}=i)=q_i, \quad \text{\it where
\quad} q_i=\frac{(1-\tau_k)^2}{\tau_k}\,i \tau_k^{i}.
\end{equation}
That is, $\eta_{n}^{(k)}$ is determined by the density function of a
$\Gamma(-\ln\tau_k,2)$-distribution and the limit distribution has the
probability generating function $q(u)=\sum_{n\ge 1}q_iu^i=
\frac{u(1-\tau_k)^2}{(1-\tau_k u)^2}$.
\end{theorem}
%%%
%%%%%%%%%%%%%%%%%%%%%%%%%%%%%%%%%%%%%%%%%%%%%%%%%%%%%%%%%%%%%%%%%%%%%%%%%
%%%
\begin{proof}
Since $g(z)=\frac{1}{1-z}$ and $h(z)=1-\frac{1}{{\bf T}_k(z)}$ have
non negative coefficients and $h(0)=0$, the composition $g(h(z))$ is
again a power series. ${\bf W}_k(z,u)=g(uh(z))$ has a singularity at
$z=\alpha_k$ and $\tau_k=1-\frac{1}{{\bf T}_k(\alpha_k)}<1=\rho_g$,
whence we are given the subcritical case. Furthermore we observe,
that regardless of the singularity arising from ${\bf T}_k(z)=0$,
the dominant singularity of $h(z)=1-\frac{1}{{\bf T}_k(z)}$ equals
the dominant singularity of ${\bf T}_k(z)$, i.e., $z=\alpha_k$.\\
%%%%%%%%%%%%%%%%%%%%%%%%%%%%%%%%%%%%%%%%%%%%%%%%%%%%%%%%%%%%%%%%%%%%%%%%%%
{\it Claim $1$.} $h(z)$ has a singular expansion at $z=\alpha_k$ and there
exists some constant $c_k>0$ such that
\begin{equation}
h(z)=
\begin{cases}
\tau_k-c_k\left(1-\frac{z}{\alpha_k}\right)^{\mu}
\ln\left(1-\frac{z}{\alpha_k}\right)(1+o(1)) &
\text{\rm for}\ k\equiv 1 \mod 2\\
\tau_k-c_k\left(1-\frac{z}{\alpha_k}\right)^{\mu}(1+o(1)) &
\text{\rm for}\ k\equiv 0 \mod 2
\end{cases}
\end{equation}
for $z\rightarrow \alpha_k$ and $\mu=(k-1)^2+\frac{k-1}{2}-1$.\\
%%%%%%%%%%%%%%%%%%%%%%%%%%%%%%%%%%%%%%%%%%%%%%%%%%%%%%%%%%%%%%%%%%%%%%%%
The proof of Claim $1$ is analogous to that of
Theorem~\ref{T:centra2}. Again, we restrict ourselves to
the case $k\equiv 1\mod 2$. ${\bf W}_{k}(z,1)=g(h(z))$ is $D$-finite
and its Taylor expansion of at $z=\alpha_k$ is given by
\begin{eqnarray*}
{\bf W}_k(z,1) &=& g(\tau_k)+(gh)'(\alpha_k)(z-\alpha_k)
+\frac{(gh)''(\alpha_k)}{2!}(z-\alpha_k)^2+\cdots\\
&=& g(\tau_k)+\frac{g'(\tau_k)\ell(\alpha_k)}{{\bf
T}_k^2(\alpha_k)}\left[\mu+\frac{\mu(\mu-1)}{2}
+\cdots\right]\left(1-\frac{z}{\alpha_k}\right)^{\mu}
\ln\left(1-\frac{z}{\alpha_k}\right)(1+o(1))\\
&=&
g(\tau_k)-c_kg'(\tau_k)\left(1-\frac{z}{\alpha_k}\right)^{\mu}
\ln\left(1-\frac{z}{\alpha_k}\right)(1+o(1)).
\end{eqnarray*}
Therefore we arrive at
\begin{equation}
[z^n]{\bf W}_k(z,1)=
c_kg'(\tau_k)\alpha_k^{-n}n^{-\mu-1}(1+o(1)).
\end{equation}
For any fixed $u\in(0,1)$ the singular expansion of ${\bf W}_k(z,u)=g(uh(z))$
at $z=\alpha_k$ is given by
\begin{align*}
{\bf W}_{k}(z,u)&=g(u\tau_k)+(g(uh))'(\alpha_k)(z-\alpha_k)
+\frac{(g(uh))''(\alpha_k)}{2!}(z-\alpha_k)^2+\cdots\\
&=g(u\tau_k)-c_ku
g'(u\tau_k)\left(1-\frac{z}{\alpha_k}\right)^{\mu}
\ln\left(1-\frac{z}{\alpha_k}\right)+(1+o(1))
\end{align*}
from which we conclude
\begin{equation}\label{E:www}
\lim_{n\rightarrow \infty}\frac{[z^n]{\bf W}_{k}(z,u)}{[z^n]{\bf
W}_{k}(z,1)}
=\frac{ug'(u\tau_k)}{g'(\tau_k)}=\frac{u(1-\tau_k)^2}{(1-\tau_k
u)^2}{\qquad \mbox{where} \quad} \tau_k=1-\frac{1}{{\bf
T}_{k}(\alpha_k)}.
\end{equation}
In view of
$[u^i]q(u)=\frac{(1-\tau_k)^2}{\tau_k}\, i \tau_k^{i}=q_i$,
Theorem~\ref{T:continuity} implies the discrete limit law
\begin{equation}
\lim_{n\rightarrow
\infty}\mathbb{P}(\eta_{n}^{(k)}=i)=
\lim_{n\rightarrow\infty}\frac{\beta_{n,i}^{(k)}}
{\beta_{n}^{(k)}}=q_i.
\end{equation}
In view of eq.~(\ref{E:gamma}), setting $r=2$ and $\lambda=-\ln
\tau_k>0$, we analogously obtain
\begin{eqnarray*}
\lim_{n\rightarrow\infty}\mathbb{P}(\eta_{n}^{(k)}=i)
&=& \frac{(1-\tau_k)^2}{\tau_k}(i\cdot \tau_k^i)
\; = \;\frac{(1-\tau_k)^2}{\tau_k}\frac{1}{(\ln \tau_k)^2}(\ln
\tau_k)^2i\cdot \tau_k^i\\
& = & \frac{(1-\tau_k)^2}{\tau_k}\frac{1}{(\ln \tau_k)^2}\ f_{-\ln\tau_k,2}(i)
\end{eqnarray*}
and Theorem~\ref{T:central} is proved.
\end{proof}

%%%
%%%%%%%%%%%%%%%%%%%%%%%%%%%%%%%%%%%%%%%%%%%%%%%%%%%%%%%%%%%%%%%%%%%%%%%%%%%%%%
%%%

%%%
%%%
%%%%%%%%%%%%%%%%%%%%%%%%%%%%%%%%%%%%%%%%%%%%%%%%%%%%%%%%%%%%%%%%%%%%%%%%
%%%

{\bf Acknowledgments.}
%%%
%%%%%%%%%%%%%%%%%%%%%%%%%%%%%%%%%%%%%%%%%%%%%%%%%%%%%%%%%%%%%%%%%%%%%%%%%%
%%%
We are grateful to W.Y.C.~Chen for stimulating discussions.
This work was supported by
the 973 Project, the PCSIRT Project of the Ministry of Education,
the Ministry of Science and Technology, and the National Science
Foundation of China.
%%%
%%%%%%%%%%%%%%%%%%%%%%%%%%%%%%%%%%%%%%%%%%%%%%%%%%%%%%%%%%%%%%%%%%%%%%%%
%%%

%%%
%%%%%%%%%%%%%%%%%%%%%%%%%%%%%%%%%%%%%%%%%%%%%%%%%%%%%%%%%%%%%%%%%%%%%%
%%%

\end{document}